\documentclass{article}
\usepackage[T1]{fontenc}
\usepackage[utf8]{inputenc}
\usepackage{ismir}
\usepackage{amsmath,cite,url}
\usepackage{graphicx}
\usepackage{color}

\title{How Much AI Is in This Track? Quantifying the Proportion of AI-Generated Stems in Hybrid Music Mixtures}
\multauthor
  {Fernando Garcia de la Cruz$^1$ \hspace{1cm} David López-Ayala$^1$ \hspace{1cm} Pablo Zinemanas$^2$}
  {{\bf Emilio Molina$^2$ \hspace{1cm} Martín Rocamora$^1$ }\\
  $^1$ Music Technology Group, Universitat Pompeu Fabra, Barcelona, Spain\\
  $^2$ BMAT Licensing S.L., Barcelona, Spain\\
  {\tt\small fergarciadlc@gmail.com}
  }

\def\authorname{F. Garcia de la Cruz, D. López-Ayala, P. Zinemanas, E. Molina, and M. Rocamora}

\usepackage[bookmarks=false,pdfauthor={\authorname},pdfsubject={\pdfsubject},hidelinks]{hyperref}

\begin{document}

\maketitle

\begin{abstract}
AI-generated music is increasingly used at the stem level, with producers integrating synthetic drums, basslines, or vocals alongside human-performed instruments. However, current AI music detection systems are binary, treating tracks as either fully AI or fully human. In this paper, we reformulate AI music detection as a regression problem on a continuous AI energy ratio, $\alpha \in [0, 1]$. We propose a methodology that leverages a multi-track music dataset to assemble mixtures of human-performed and AI-reconstructed stems (obtained using a neural audio codec) with known proportions of each content type. Using this approach, we first show that a CNN-based model trained on fully AI-generated or human-performed tracks, which achieves $>99\%$ accuracy as a binary detector, when faced with mixed content, yields an output that rises with the AI stems' energy contribution, acting as a noisy and miscalibrated estimator. Our analysis of the influence of different stems shows that detection sensitivity depends on the instrument and reflects its frequency content: drums and guitar carry strong codec-artifact signatures, while vocals and bass are less detectable. Based on these insights, we train a similar CNN-based model for regression of $\alpha$, achieving $\text{MAE} = 0.076$ and $R^2 = 0.85$ on held-out mixtures from the same pipeline. These results suggest that the regression formulation is an initial promising step towards AI-music detection in realistic music production workflows.
\end{abstract}

\section{Introduction}\label{sec:introduction}

The prevalence of AI-generated content is growing rapidly. Industry reports estimate that fully AI-generated tracks now account for more than 50\% of daily uploads on major streaming platforms \cite{Wendel2026DeezerAIShare}.
Industry studies also report that 97\% of listeners cannot reliably distinguish fully AI-generated tracks from human-made recordings \cite{Wendel2025DeezerStudy}. This raises concerns about transparency and fair remuneration for artists and highlights the need for automated detection tools that are also aligned with disclosure obligations, such as those in Article 50 of the EU AI Act \cite{Serra2026AIActMarking}.

Current AI music detection systems achieve near-perfect accuracy (>99\%) when classifying complete tracks as fully AI-generated or human-produced~\cite{Afchar2025AIGeneratedMusic,Rahman2025SONICS}. These detectors exploit architectural artifacts: the characteristic spectral peaks introduced by transposed convolution layers in neural audio codec decoders~\cite{Afchar2025FourierExplanation}. Yet, it is unclear how this binary paradigm applies when AI-generated stems are combined with human performances on a single track \cite{Pons2025MusicAITrends}.

In this work, we reformulate AI music detection as a regression problem on a continuous AI energy ratio $\alpha \in [0, 1]$, representing the proportion of a mix's total acoustic energy contributed by AI-generated stems. Studying this reformulation requires controlled mixtures of human-performed and AI-generated stems with known $\alpha$ values. 

To enable this, we propose a methodology for constructing hybrid mixtures from multi-track music datasets based on three components: (i) encode--decode reconstruction of human-performed stems through a neural audio codec, which imprints the same architectural artifacts that binary detectors learn to detect while preserving the original musical content; (ii) combinatorial stem replacement, which for a track with $n$ stems produces all $2^n$ real/AI configurations, ensuring dense sampling of $\alpha$; and (iii) evaluation at the level of the AI energy ratio $\alpha$, which correctly weights each stem by its acoustic contribution.

We apply this methodology using the MoisesDB multi-track dataset \cite{Pereira2023MoisesDB} and the EnCodec neural audio codec \cite{Defossez2022Encodec}, producing controlled mixtures across 240 professionally produced tracks. Our experiments reveal that a state-of-the-art binary detector \cite{Afchar2025AIGeneratedMusic}, which achieves $>99\%$ accuracy on pure tracks, produces an output score that increases with $\alpha$, effectively acting as a noisy and miscalibrated estimator of the AI energy ratio. Detailed analysis of the contribution of the different stems to the output indicates that the detection sensitivity is frequency-dependent: while drums and guitars are more reliably identified, vocals are only partially detected, and bass remains largely undetected. Building on these findings, we train a CNN regressor using the same architecture to estimate $\alpha$ directly, achieving a Mean Absolute Error (MAE) of $0.076$ ($R^2 = 0.85$) on held-out mixtures from the same reconstruction pipeline.

The AI stems in our study are codec reconstructions of real recordings, not generator output, so what we measure is the proportion of codec-generated energy under controlled conditions (\secref{sec:limitations}). The binary detector is reproduced from Afchar et al.~\cite{Afchar2025AIGeneratedMusic} as a diagnostic probe; our contributions are the reformulation, the methodology and released library, the characterization of binary-detector behavior on hybrid mixtures, and a regression baseline.

\section{Related Work}\label{sec:related_work}

Neural audio codecs \cite{Defossez2022Encodec,Kumar2023RVQGAN} enable high-fidelity audio compression into discrete token sequences, and audio and music generation systems \cite{Borsos2023AudioLM, Copet2023MusicGen} leverage these representations to generate music autoregressively from text or audio prompts. Other systems generate in the continuous latent space of an audio autoencoder. Stable Audio \cite{Evans2024StableAudio} synthesizes full tracks from text with latent diffusion, and Diff-A-Riff \cite{Nistal2024DiffARiff,Nistal2024ImprovingDiffARiff} uses the same paradigm to generate one instrument stem at a time, conditioned on a reference mix through audio or text embeddings, while LiveBand \cite{Pasini2026LiveBand} generates accompaniment causally with an adversarially trained transformer. The latter two produce hybrid tracks directly. Whatever the conditioning, all of these systems synthesize the waveform through a learned decoder built from transposed convolution layers. These layers introduce periodic spectral artifacts \cite{Odena2016Checkerboard,Pons2021UpsamplingArtifacts}, and current AI-music detectors learn to exploit them \cite{Afchar2025FourierExplanation}.

Early work on AI music detection focused on binary classification of complete tracks. Afchar et al.~\cite{Afchar2025AIGeneratedMusic,Afchar2025FourierExplanation} proposed a CNN-based detector that reaches 99.8\% accuracy on controlled datasets by exploiting codec-induced spectral artifacts. Rahman et al.~\cite{Rahman2025SONICS} proposed \textit{SpecTTTra}, a transformer with spectro-temporal tokenization that reaches an F1 score of 97\% on full-track detection. However, these detectors face important limitations. Generalization difficulties across generators were already noted by Afchar et al., and Cros Vila et al.~\cite{Vila2025AIMusicArmsRace} further demonstrated that binary detectors exhibit strong platform-specific biases: a detector trained on music from one generation platform fails to generalize to another. What these detectors learn is therefore specific to a generator, not a universal signature of AI-generated audio. Afchar et al.~\cite{Afchar2025AIGeneratedMusic} also showed that simple audio transformations, such as pitch shifting or re-encoding, significantly degrade detection accuracy, exposing vulnerability to post-processing. Detection also degrades on broadcast audio, where music is often short and mixed with speech; F1 scores fall below 60\% \cite{LopezAyala2026Broadcast}.

A similar redefinition of the detection of AI-generated content has already taken place in adjacent fields. In visual deepfake detection, recent work reformulates the task as multi-label classification over individual facial components or predicts pixel-level masks that localize manipulated regions within video frames \cite{Deng2024MultiLabelDeepfake, Xu2025SelfBlending}. 

While existing detectors achieve high accuracy (>99\%) on binary classification of fully AI-generated tracks, to our knowledge no prior work has examined detection when AI-generated content is mixed with human-created content at varying proportions. This setting is increasingly relevant as professional production integrates AI-generated components with human performances \cite{Pons2025MusicAITrends}, a practice that the stem- and accompaniment-generation models discussed above directly support. Yet detectors trained under the assumption of either fully AI or fully human music can exhibit degraded performance on mixed content (\secref{sec:experiments}). Motivated by this gap, and supported by the shift toward partial detection in adjacent fields, we study proportional AI content detection at the stem level and propose a regression baseline for the task.

\section{Problem Formulation}\label{sec:problem_formulation}

\subsection{Binary AI-music Detection}

In prior work, AI music detection has been formulated as a binary classification problem~\cite{Afchar2025AIGeneratedMusic}.
Given an audio signal $\mathbf{x} \in \mathbf{R}^{T \times C}$ (where $T$ is the number of samples
and $C$ is the number of channels), the task is to predict $y \in \{0, 1\}$,
where $y = 0$ indicates human-produced music and $y = 1$ indicates AI-generated music.

The detectors identify artifacts in the form of spectral peaks introduced by the transposed convolution layers of the decoder, the audio counterpart of the \textit{checkerboard
artifacts} of image generation~\cite{Odena2016Checkerboard}. Afchar et al.~\cite{Afchar2025FourierExplanation} call this signature an \emph{artifact fingerprint}, and in their released code, a \emph{fakeprint}, the term we use here. The peaks appear at
$f_{\text{artifact}} = f_s / (2 \times \text{stride})$, where $f_s$ is the sampling rate. Because the strides are fixed by the architecture, the peaks appear at the same frequencies in all audio the decoder generates.

\subsection{Mixed-Content Production Scenarios}
Assuming a linear mixture, a track is the sum of $n$ stems, $x_{\text{mix}} = \sum_{i=1}^{n} s_i$,
where each stem $s_i$ is either human-produced ($s_i \in H$) or AI-generated 
($s_i \in A$). 
To quantify the AI content of a mix, we define the \emph{AI energy ratio}

\begin{equation}
    \alpha = \frac{\sum_{s_i \in A} \mathrm{RMS}(s_i)^2}{\sum_{i=1}^{n} \mathrm{RMS}(s_i)^2},
    \label{eq:alpha}
\end{equation}

\noindent where $\mathrm{RMS}(\cdot)$ denotes the root-mean-square waveform amplitude values of a stem and $\alpha$ 
is the fraction of total signal energy contributed by AI-generated stems.

Mixing alters artifact detectability. The spectral peaks are present only in the AI stems, so combining them with human content in overlapping frequency bands reduces their prominence relative to the surrounding spectrum. 

\subsection{Proportional Detection via Regression}

We reformulate AI music detection as a regression problem. Given a
mixed audio signal $x_{\text{mix}}$, the task is to predict
$\hat{\alpha} = f_\theta(x_{\text{mix}}) \in [0, 1]$, where $\hat{\alpha}$
is the estimated AI energy ratio and $f_\theta$ is a learned function
parameterized by $\theta$.

We report mean absolute error (MAE) and root mean squared error (RMSE) in the
units of $\alpha$, the coefficient of determination ($R^2$), and Pearson's $r$. A proportional output such as
``approximately $X\%$ AI-generated'' carries more information than a binary label and better
reflects hybrid production workflows.

\subsection{Key Challenges}

This reformulation introduces two challenges beyond binary classification. Regression must
distinguish arbitrary values of $\alpha$, and the same $\alpha$ can arise from different stem
combinations (50\% AI vocals versus 50\% AI drums), so detectability depends on \emph{how much}
energy the AI stems contribute and on \emph{where} in the spectrum that energy lies.

\section{Methodology}\label{sec:methodology}
We introduce a methodology for constructing controlled hybrid mixtures of
human-performed and AI-generated stems with a known AI energy ratio
$\alpha$, as defined in \eqnref{eq:alpha}. Note that the AI-generated stems in our study are codec reconstructions of real recordings.

The methodology is guided by three principles:
(1) \emph{stem-level granularity}, reflecting the modular structure of modern
production; (2) \emph{artifact isolation through reconstruction}, which reduces
content confounds between AI and human conditions; and (3)
\emph{combinatorial completeness}, which provides dense coverage of $\alpha$
over the full $[0,1]$ range. The methodology is agnostic to both the source corpus
(any multi-track dataset with isolated stems can serve as input)
and the codec used for reconstruction.

\subsection{AI Reconstruction Pipeline}\label{subsec:pipeline}

We build AI stems by encoding and decoding existing human-performed stems
through a neural audio codec. This encode--decode process introduces the characteristic spectral
artifacts of the decoder's transposed convolution layers \cite{Afchar2025FourierExplanation}
while preserving the original musical content. This isolates the detection problem to
architectural artifacts, eliminating confounds from differences in musical style or
content between the AI and human conditions. The process is deterministic and therefore reproducible.
The methodology is also codec-agnostic; any neural codec exposing encode and decode operations can be substituted. 

We instantiate the pipeline with EnCodec \cite{Defossez2022Encodec}, 
reconstructing stems at a single bitrate of 3~kbps. 
%
%
When reproducing the evaluation conditions in~\cite{Afchar2025AIGeneratedMusic}, the reconstructed stems are stored as MP3 at 256~kbps, 44.1~kHz, 
so that any observed detection failure on mixed content is attributable to the mixing process and not to a format mismatch.

\subsection{Combinatorial Mix Generation}\label{subsec:mix_generation}

Given a track with $n$ stems indexed by $I = \{0, 1, \ldots, n - 1\}$, we 
define a mix specification by the subset $S \subseteq I$ of stems replaced 
with their AI-reconstructed versions. The resulting mix is
\begin{equation}
    x_{\text{mix}}(S) = \sum_{i \in I \setminus S} s_i^{\text{real}} + \sum_{i \in S} s_i^{\text{AI}},
    \label{eq:mix_subset}
\end{equation}
where $s_i^{\text{real}}$ and $s_i^{\text{AI}}$ denote the authentic and 
AI-reconstructed versions of stem $i$. Stems are summed with equal weights, 
preserving the original energy relationships defined by the source mix.

We organize mixes into sets indexed by the number of AI-reconstructed stems:
\begin{equation}
    U_k = \left\{ x_{\text{mix}}(S) : S \subseteq I,\; |S| = k \right\}, \quad k = 0, 1, \ldots, n.
\end{equation}
Each set $U_k$ contains $\binom{n}{k}$ mixes, for a total of $2^n$ mixes 
per track. The ratio $k/n$, the \emph{stem fraction}, counts \emph{how many} stems are replaced, whereas the target variable $\alpha$ (\eqnref{eq:alpha}) weights each stem by its energy. 

Sets $U_0$ (all real) and $U_n$ (all AI) correspond to the binary detection setting; the
intermediate sets $U_1, \ldots, U_{n-1}$ contain the mixed-content cases. Within a set, mixes
share the stem fraction $k/n$ but span a range of $\alpha$ values and differ in which instruments
are AI. This last property is what enables the instrument-level analysis of
\secref{subsec:instrument_sensitivity}.

\subsection{Multi-track audio dataset}\label{subsec:instantiation}
The methodology applies to any multi-track dataset providing isolated stems, 
including those developed for source separation (e.g., MUSDB18 \cite{Rafii2017MUSDB18}, Slakh \cite{Manilow2019Slakh}). 
We instantiate the methodology on MoisesDB \cite{Pereira2023MoisesDB}, a professionally 
mixed multi-track dataset providing high-quality stem isolations across 
diverse genres. We use 240 tracks, with stem counts ranging from 2 to 10 (most tracks contain 4--7 stems) 
categorized by instrument type (vocals, drums, bass, guitar, and other 
instruments such as keyboards, strings, and synthesizers). The variable stem count is a feature of the design: tracks with different $n$ produce different sets of 
achievable $\alpha$ values, and aggregated across the corpus, these provide dense 
coverage of the full $[0,1]$ range, with most mixes concentrated around mid-range values. In total, this yields 21{,}212 unique mixes across the 240 tracks.

We release a reproducible Python library\footnote{ARIA (Authentic vs. Reconstructed Isolated Audio). Methodology and training code available at \url{https://fergarciadlc.github.io/ismir26-how-much-ai/}.} that constructs mixtures from a local MoisesDB installation.

\section{Experiments and Results}\label{sec:experiments}

We apply the methodology of \secref{sec:methodology} to construct hybrid mixtures and compare two models on them: the reproduced binary detector, and a regression model with the same backbone trained on $\alpha$.

\subsection{Experimental Setup}\label{subsec:experimental_setup}

\textbf{Models}. Following \cite{Afchar2025AIGeneratedMusic}, we implement a CNN classifier: a six-block convolutional backbone operating on magnitude spectrograms (STFT with 2048-sample window, 512-sample hop, frequencies above 16\,kHz discarded), followed by global average pooling and a dense head. We train two variants of this architecture that differ only in their output layer and loss. The binary detector uses a sigmoid-activated classification head trained with binary cross-entropy on the original $\{0, 1\}$ labels. 
The regression model replaces the classification head with a single-unit, sigmoid-activated regression head trained with mean squared error against the AI energy ratio $\alpha$ (\eqnref{eq:alpha}). 
At inference, both models produce per-window scores $\hat{y} \in [0,1]$, which are averaged across non-overlapping windows to obtain a track-level score.

\textbf{Training.} The binary detector is trained on FMA-Medium \cite{Defferrard2017FMA_dataset} and its EnCodec-reconstructed counterpart at 3\,kbps, following the original methodology by Afchar et al. \cite{Afchar2025AIGeneratedMusic}. The regression model is trained on the hybrid mixtures produced by our methodology applied to MoisesDB at the 3\,kbps EnCodec condition (\secref{subsec:instantiation}), with $\alpha$ computed per 5-second window as the regression target. 

We initially trained both models on 1-second windows, matching prior work \cite{Afchar2025AIGeneratedMusic}. The regression model performed better with 5-second windows, particularly near the boundaries of the $\alpha$ range where short windows produced noisier estimates, so we retrained the binary detector at 5-second windows as well and report the matched comparison throughout. \secref{subsec:regression_baseline} reports both window lengths.
Train, validation, and test splits are performed at the track level, stratified by stem count, so that all mixes and all windows from a given track land in the same partition, preventing content and temporal-correlation leakage. The 240 tracks are split 80/10/10, leaving 24 held-out tracks and 1{,}792 test mixtures.

\textbf{Reproduction validation.} The binary detector achieves 99.97\% accuracy and 99.98\% F1 on a held-out FMA evaluation set, closely matching \cite{Afchar2025AIGeneratedMusic}. On the pure endpoints \(U_0\) and \(U_n\) of our mixture corpus, the detector assigns scores near 0 and 1, respectively, confirming that the reconstruction pipeline preserves the codec artifacts on which the binary classifier relies.

\textbf{Evaluation.} We run inference on the held-out test split and report track-level scores, obtained by averaging the per-window scores of each mixture; the instrument-level analysis of \secref{subsec:instrument_sensitivity} uses window scores directly.

\subsection{Energy-Driven Detection Response}\label{subsec:energy_detection_response}

\begin{figure}[t]
  \centering
  \includegraphics[width=\linewidth]{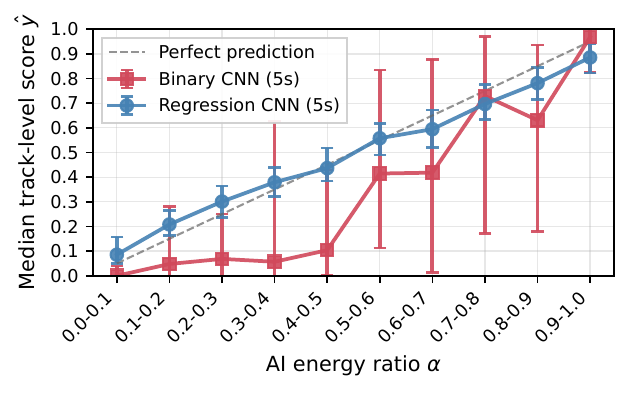}
  \caption{Median $\hat{y}$ as a function
  of the AI energy ratio $\alpha$ for the binary and regression models, both at 5-second windows; error bars show the interquartile range.}
  \label{fig:median_pred_alpha}
\end{figure}

We first examine how each model's output $\hat{y}$ depends on the AI energy ratio $\alpha$ (Figure~\ref{fig:median_pred_alpha}, medians binned in 10\% increments, both models at 5-second windows).

The binary detector's output increases with $\alpha$, but the increase is not linear: its median stays near $0.05$ for $\alpha < 0.5$ and then rises to $\sim 0.97$ for $\alpha > 0.9$. The detector was trained only on the endpoints $\alpha \in \{0, 1\}$ and has never seen mixed content, so this response shows that codec artifacts in hybrid mixtures carry proportional information about AI content, which the binary classifier picks up incidentally as a noisy and miscalibrated estimator of $\alpha$.

The regression model, trained directly on $\alpha$, follows the diagonal across the full range, while the binary detector underestimates AI content in the low and mid range: at $\alpha \in [0.4, 0.5]$ its median is $0.10$, against $0.44$ for the regression model. Part of this gap comes from window length (\secref{subsec:regression_baseline}). The proportional signal is therefore present in mixed content, and the training objective determines how well it is recovered.

\subsection{Instrument-Level Detection Sensitivity}\label{subsec:instrument_sensitivity}

\begin{figure}[t]
  \centering
  \includegraphics[width=\linewidth]{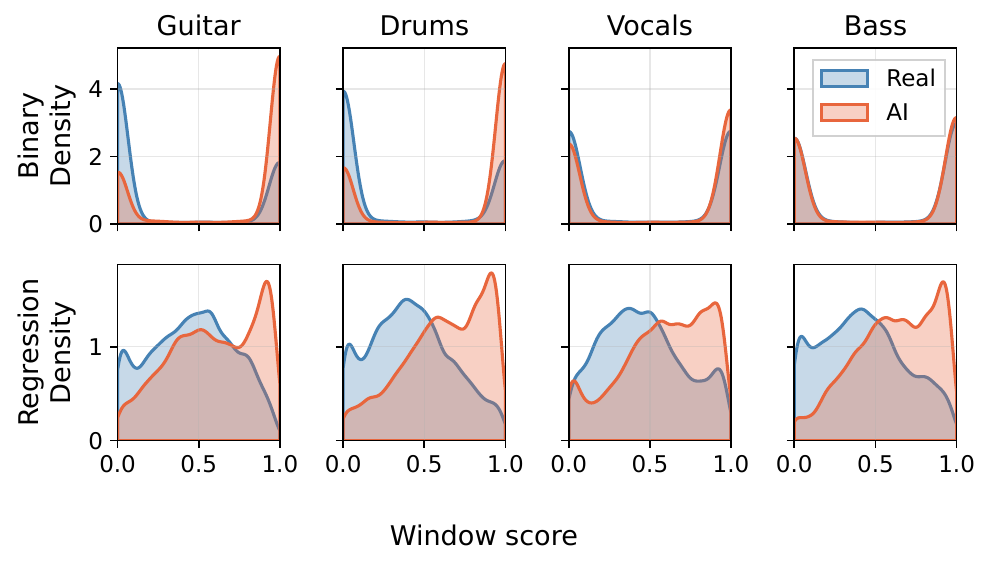}
  \caption{
  Kernel density estimates of $\hat{y}$ for each instrument under real and AI conditions, for the binary detector (top row) and regression model (bottom row), both at 1-second windows.
  }
  \label{fig:density_cdf}
\end{figure}

A natural place to look for the source of detection sensitivity is the stem level, the unit at which AI-generated content is integrated into modern productions. We isolate each instrument's contribution to detection. We select tracks in our test corpus whose stems are exactly $\{$vocals, drums, bass, guitar$\}$, the four most common stem types in the corpus (instantiated from~\cite{Pereira2023MoisesDB}). Each of these appears individually in over 90\% of tracks, whereas the next most common type, piano, is present in only 46\%. For each selected track we generate all 16 possible AI/real configurations, excluding the all-real case. For each instrument, we compare $\hat{y}$ when that stem is real versus AI-generated, marginalizing over the other three stems. Figure~\ref{fig:density_cdf} shows the resulting score distributions by window for both models, at 1-second windows. This is the setting in which the binary detector tracks $\alpha$ most closely (Table~\ref{tab:regression_results}), and therefore the most favorable one for it.

For the binary detector (top row), drums and guitar separate most clearly, with the AI distribution concentrated near 1 and the real distribution near 0; whereas the distributions for vocals and bass substantially overlap. 
Sensitivity therefore depends on the instrument, and \secref{subsec:fakeprint_analysis} traces this to where each stem's energy lies in the spectrum. 
Notably, vocals and bass become partially separable under regression (bottom row), indicating that the regressor can extract proportional information from a weaker fakeprint signal even when the binary detector cannot.

\subsection{Fakeprint Analysis}\label{subsec:fakeprint_analysis}

\begin{figure}[t]
  \centering
  \includegraphics[width=\linewidth]{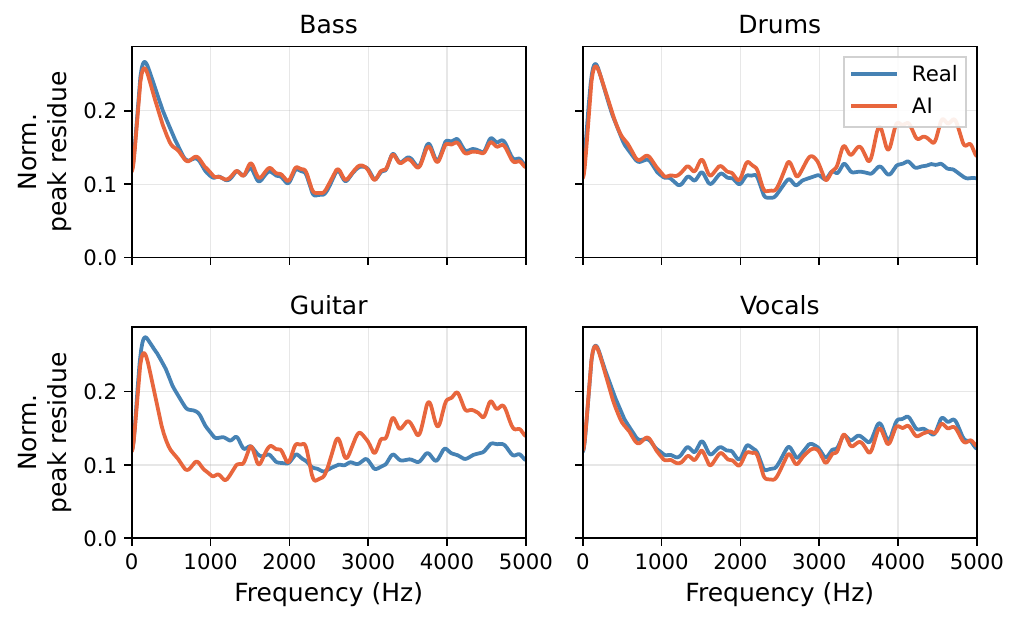}
  \caption{Average fakeprint curves (Gaussian-smoothed) for Real and AI stems across bass, drums, guitar, and vocals over the $[0, 5]$\,kHz band, highlighting stem-specific spectral differences between real and generated content.}
  \label{fig:fakeprints}
\end{figure}

To verify that the per-instrument differences in detection observed in \secref{subsec:instrument_sensitivity} reflect the underlying artifact distribution rather than other detector's factors, we compute the fakeprints directly from the stems. Following Afchar et al.~\cite{Afchar2025FourierExplanation}, we obtain a fakeprint by time-averaging the log-power spectrum of a stem and subtracting the local minima over sliding frequency windows, keeping only the positive residual peaks. Theory predicts that AI-generated content exhibits peaks at frequencies determined by the strides of the codec's transposed convolution layers~\cite{Afchar2025FourierExplanation}; whereas real recordings do not present these artifacts. We follow the original procedure but restrict the bandwidth to $[0, 5]$\,kHz to capture the lower half of the audible range where instrument fundamentals and lower harmonics live; at this bandwidth and after Gaussian smoothing, the curves capture broad spectral envelope differences between real and AI conditions rather than the narrow stride-determined peaks predicted by theory. Following the same marginalization strategy as \secref{subsec:instrument_sensitivity}, we pool fakeprints from all mixes in which a given stem is AI-generated versus real and average within each group. Figure~\ref{fig:fakeprints} shows the average fakeprints for the four different types of stems.

The fakeprint differences exhibit the following patterns. Bass shows essentially no separation between real and AI conditions, whereas vocals show very subtle differences. 
In contrast, drums show clear separation from $\sim 2500$\,Hz onward, with the AI curve lifted above the real curve. Guitar shows a distinctive crossover: the AI curve sits below the real curve in the low-mid range ($\sim 500$--1500\,Hz) and substantially above it from $\sim 2500$\,Hz onward, suggesting that AI-reconstructed guitar both attenuates low-mid spectral structure and introduces high-frequency artifact peaks. 
These results align with 
the binary detector's instrument sensitivity reported in \secref{subsec:instrument_sensitivity}, indicating that the variation is rooted in the codec's artifact distribution: bass and vocal evade detection because the codec leaves a lower detectable artifact signature in their content, while drums and guitar carry strong signatures that the detector can identify.


\subsection{Regression Baseline}\label{subsec:regression_baseline}

Section~\ref{subsec:experimental_setup} describes the training setup; here we report performance on the held-out test split.

\textbf{Baseline.} As an implicit baseline we use the binary detector directly as an $\alpha$ estimator, treating its fake probability $\hat{y}$ as a prediction of $\alpha$. This is an unfair comparison, since the binary detector was not trained for this objective, but it quantifies how much information about $\alpha$ a binary classifier captures incidentally, which is exactly the question raised by \secref{subsec:energy_detection_response}.

\begin{figure}[t]
  \centering
  \includegraphics[width=\linewidth]{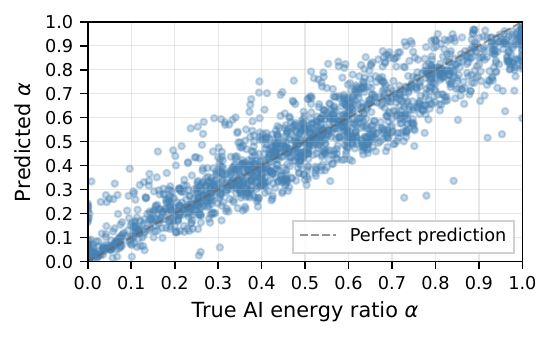}
  \caption{
  Predicted versus true AI energy ratio $\alpha$ for the regression model, averaged across the windows of each mixture, on the held-out test split.
  }
  \label{fig:scatter}
\end{figure}

\textbf{Results.} Table~\ref{tab:regression_results} reports performance on the test split. At the matched 5-second window, the regression model achieves MAE = $0.076$, RMSE = $0.102$, $R^2 = 0.85$, and Pearson $r = 0.93$. The binary detector's implicit estimate reaches MAE = $0.289$ and $R^2 = -0.85$. A negative $R^2$ means the estimate is less accurate than always predicting the dataset mean, even though the same detector classifies the pure endpoints with $>99\%$ accuracy. Training directly on $\alpha$ therefore gives a large gain over reusing a binary score. Figure~\ref{fig:scatter} shows predicted vs.\ true $\alpha$ averaged across the windows at the track level.

\textbf{Window length.} Table~\ref{tab:regression_results} also reports both models at 1-second windows, and the two objectives respond to window length in opposite directions. The regression model improves with longer windows (MAE $0.076$ vs.\ $0.104$, $R^2$ $0.85$ vs.\ $0.78$), since a continuous quantity is estimated more stably over more content; this motivates the 5-second choice. The binary detector, which is never trained on $\alpha$, gets worse ($R^2$ $-0.19$ vs.\ $-0.85$), since averaging over a longer window blurs the weak proportional signal it captures. The regression model is more accurate at both window lengths, and the matched 5-second setting is the one we compare on.

\textbf{Boundary compression.} The regression model exhibits mild compression near the boundaries: predictions floor at $\sim 0.05$--$0.10$ for fully real mixtures and ceiling at $\sim 0.85$--$0.95$ for fully AI mixtures, explaining the convergence with the binary detector at $\alpha = 1$ noted in \secref{subsec:energy_detection_response}. This is consistent with bounded sigmoid regression under MSE loss. Absolute errors nevertheless remain low at the boundaries; the largest errors occur in the mid range, where mixtures admit the most varied stem compositions and the analyses of \secref{subsec:instrument_sensitivity} and \secref{subsec:fakeprint_analysis} predict the greatest variance in detectability.

\begin{table}[t]
  \centering
  \resizebox{\columnwidth}{!}{%
  \begin{tabular}{lcccc}
    \hline
    \textbf{Model} & \textbf{MAE} & \textbf{RMSE} & \textbf{R$^2$} & \textbf{Pearson}\\
    \hline
    Binary CNN, 1\,s ($\alpha$-est.) & 0.228 & 0.289 & $-$0.193 & 0.617 \\
    Binary CNN, 5\,s ($\alpha$-est.) & 0.289 & 0.360 & $-$0.851 & 0.505 \\
    Regression CNN, 1\,s & 0.104 & 0.124 & 0.782 & 0.892 \\
    Regression CNN, 5\,s & \textbf{0.076} & \textbf{0.102} & \textbf{0.854} & \textbf{0.925} \\
    \hline
  \end{tabular}%
  }
  \caption{Proportional-detection performance on the held-out test split, computed on per-mixture scores, each averaged across the windows of the mixture, at 1- and 5-second windows. The binary detector's raw fake probability is used as an implicit $\alpha$ estimate.}
  \label{tab:regression_results}
\end{table}

\section{Limitations}\label{sec:limitations}

Our empirical findings are based on a single neural audio codec (EnCodec~\cite{Defossez2022Encodec}) at a single bitrate (3\,kbps). The AI stems are codec reconstructions 
which isolate codec artifacts from content confounds and enable controlled, reproducible experiments, but bound the scope of our claims. Whether these findings transfer to other codecs (e.g., DAC~\cite{Kumar2023RVQGAN}) or to commercial generators such as Suno and Udio, whose artifacts may lie at different frequencies, remains untested. That said, the artifacts are a property of the decoder architecture~\cite{Afchar2025FourierExplanation}, and our methodology does not depend on which architecture produces them. 

Our mixtures are also constructed without the post-processing typical of professional production workflows (EQ, compression, level balancing, creative editing of AI stems), which could alter both detection accuracy and the effectiveness of the AI energy ratio estimation. Accordingly, we view this work as a controlled methodological foundation rather than a deployment-ready detector. Translating these findings into production tools requires corpora in which human-performed stems are paired with stems from actual music generators, for which stem-level access is currently limited. Two routes could supply them: collaboration with music platforms holding stem-level catalogs, and stem-conditioned generative models \cite{Nistal2024DiffARiff,Pasini2026LiveBand}, which generate stems that fit existing human-performed material. The methodology and library we release are codec- and corpus-agnostic and can be instantiated on such data as it becomes available.

\section{Conclusion}\label{sec:conclusion}
In this work, we address AI-music detection on mixed content in a controlled setting by reformulating the task as regression on a continuous AI energy ratio $\alpha$ and proposing a codec- and corpus-agnostic methodology for constructing hybrid mixtures with known $\alpha$ from any multi-track corpus. Applying this methodology to MoisesDB and EnCodec, we show that codec artifacts in hybrid mixtures carry proportional information about $\alpha$: a binary detector trained only on $\alpha \in \{0, 1\}$ captures this signal incidentally as a noisy estimator, whereas a regression model with the same backbone trained directly on $\alpha$ extracts it substantially more accurately (MAE $= 0.076$, $R^2 = 0.85$). Detection sensitivity depends on the stem: some carry stronger codec-artifact signatures than others, in line with their frequency content, as the fakeprint analysis confirms.

The proportional-detection formulation offers a more informative target than binary classification for the realistic case of hybrid music production, and the methodology can be instantiated with other codecs and with stems from commercial AI generators. Because sensitivity depends on the stem, instrument-aware detection is a natural next step: applying detection to separated stems before aggregation, or conditioning on stem identity during training. Both models also produce per-window scores, so reporting them over time gives a route to temporal localization; detecting generated material within a stem, such as a single synthetic drum hit, would require mixtures built at finer granularity than the stem. The methodology and accompanying library are released as a foundation for continued research on hybrid AI-music detection.

\section{Acknowledgments}\label{sec:acknowledgments}
This work is supported by the ``Cátedra IA y Música'' project (TSI-100929-2023-1), funded by the Secretaría de Estado de Digitalización e Inteligencia Artificial, the European Union-Next Generation EU funds and BMAT Music Innovators. And by the ``IMPA'' project (PID2023-152250OB-I00) funded by MCIU/AEI/10.13039/501100011033/FEDER, UE.

\bibliography{ISMIRtemplate}
\end{document}